\documentstyle[11pt]{article}
\title{Dynamics of Relativistic Interacting Gases~:\\
From a Kinetic to a Fluid Description}
\author{Jean-Philippe Uzan \\
D\'epartement d'Astrophysique Relativiste et de Cosmologie\\
UPR 176 du Centre National de la Recherche Scientifique\\
Observatoire de Paris, 92195 Meudon, France}
\def\be{\begin{equation}}
\def\ee{\end{equation}}
\def\bt{\begin{tabular}}
\def\et{\end{tabular}}
\def\lp{\left(}
\def\rp{\right)}
\def\lb{\left\lbrace}
\def\rb{\right\rbrace}
\def\bea{\begin{eqnarray}}
\def\eea{\end{eqnarray}}
\def\P{{\cal P}}
\def\h{{\cal H}}
\def\D{\Delta}
\def\T{{\cal T}}
\def\Pm{{\cal P}_m}
\def\p{{\bf p}}
\def\q{{\bf q}}
\def\k{{\bf k}}
\def\u{{\bf u}}
\def\x{{\bf x}}
\def\pp{{\bf p'}}
\def\qp{{\bf q'}}
\def\kp{{\bf k'}}
\begin{document}
\maketitle
%-------------------------------------------------------------------------- 
\begin{abstract}
%-------------------------------------------------------------------------- 

Starting from a microscopic approach,
we develop a covariant formalism to
describe a set of interacting gases. For that purpose,
we model the collision term entering the Boltzmann equation
for a class of interactions and then integrate this equation to
obtain an effective macroscopic description.
This formalism will be useful
to study the cosmic microwave background 
non-perturbatively in inhomogeneous cosmologies.
It should also be useful for the study of the dynamics of the early universe
and can be applied, if one considers fluids of galaxies, to the
study of structure formation.
\end{abstract}

{\bf Pacs~: 98.80.Hw, 04.20.Cv}

%------------------------------------------------------------------------- 
\section{Introduction}\label{par1}
%------------------------------------------------------------------------- 

The formation and evolution of 
cosmological perturbations, leading for instance to
the cosmic
microwave background fluctuations, are usually studied
in the framework of cosmological perturbations about
a Friedmann-Lema\^ \i tre spacetime \cite{HS}. In such a framework, the matter
is described by a set of decoupled species including
cold dark matter, neutrinos and a baryon-photon fluid (since
one must consider the coupling between these two species
via Compton scattering). In the latter case, the interaction between
the two coupled components (i.e. photons and baryons) is studied
via the linearised Boltzmann equation.
A first generalisation that comes to mind is to
develop a framework to model this interaction in 
a covariant way, i.e. without specifying the spacetime geometry.
Such an approach will be useful for a study of the
cosmic microwave background in inhomogeneous cosmologies \cite{MCALLUM,KRA}.
Earlier in the history of the
universe all these fluids were coupled through scattering
and/or annihilation process \cite{WEINBERG}. This leads to
the second generalisation which is to model
this coupling in a very general way
for a wide class of interactions. Such a formalism
will then be adapted to study the dynamics during the early
universe, independently of the spacetime geometry. 
Moreover, it can be applied to structure formation,
since it can describe fluids of galaxies.

The Einstein equations and the Bianchi
identities lead to the fact that the 
total energy-momentum tensor of matter is conserved
\be 
\nabla_\mu T^{\mu\nu}=0.
\ee
When the matter content of the universe is a single perfect fluid,
this equation splits into two equations, the Euler equation and the
matter conservation equation. The latter can be reset as
\be 
\nabla_\mu n^\mu=0,
\ee
meaning that the number flux vector, $n^\mu$, is conserved.  

If one now considers a system composed of an
arbitrary number of fluids, this property applies
to the ``global fluid" and to each fluid only if they are
interacting by gravitational interaction 
(i.e. if they are decoupled). If we want to study the evolution of
one particular fluid of a system of interacting fluids,
we have to take into account a force and the preceeding
law can be rewritten as
\be 
\nabla_\mu T^{\mu\nu}_i=F^\nu_i.\label{F3}
\ee  
Furthermore, if the collisions are not elastic, then the
total number of each species will not be conserved and we
will have to take into account a source term
\be 
\nabla_\mu n^\mu_i=\epsilon_i.\label{F4}
\ee

Such a splitting of the global energy-momentum tensor 
($T^{\mu\nu}=\sum T^{\mu\nu}_i$) and 
of the number flux vector ($n^\mu=\sum n_i^\mu$)
is not straightforward. In fact,
for a general multi-constituent fluid \cite{CARTER}
 the lagrangian of the
system, ${\cal L}$ say, will not split in the sum of individual 
lagrangians describing 
the dynamics of each fluid. Since the energy-momentum tensor
is related to this lagrangian via
$$T^{\mu\nu}=\frac{2}{\sqrt{-g}}\frac{\delta{\cal L}
\sqrt{-g}}{\delta g_{\mu\nu}},$$
one cannot define the energy-momentum tensor of a single
component of the fluid system. However, as long as we are
dealing with gases, such a splitting will be possible. Indeed, 
each fluid is then conveniently
described by its distribution function, $f_i$ say, from
which one can define an energy-momentum tensor (see \&\ref{par2}
and \&\ref{par3} for the complete definitions) as
$$T^{\mu\nu}_i=\int p^\mu p^\nu f_i(\x,\p) \pi_+(\p).$$
This standard kinetic theory approach \cite{EHLERS} will
be valid if the particles interact weakly, i.e. for
systems which are not too dense. As explained in the
section \ref{spar64}, such a description is adapted to cosmology.

A way to model the interacting forces, $F^\mu_i$, and the source terms, 
$\epsilon_i$, is to relate them to the collision term entering the
equation of evolution of the distribution function. We will
then start from the kinetic theory, try to give 
a general form of the collision term for a class
of interactions, compute the forces and the source terms,
and end up with
a macroscopic description of the dynamics of interacting fluids.
It will be  non perturbative, in the sense that
we will not linearise the metric, and will not assume anything on
the form of the fluid (e.g. perfect fluid). 

Both the kinetic theory and the theory of fluid dynamics have been
studied in the context of general relativity.  On the one hand, a
general covariant formulation of kinetic theory in general relativity
was first developped by Tauber and Weinberg \cite{TAUBER} and
independentely by Chernikov \cite{CHERNIKOV} (see also Marle
\cite{MARLE} and Israel and Stewart \cite{IS2}). These authors were
mostly concerned with relativistic generalisations of classical gas
theory (proof of the H-theorem, equilibrium configurations etc.).
Many authors have also studied the coupled Einstein-Boltzmann equation
and solved it \cite{EHLERS2,MISNER} in some cases. It is not our goal
to solve such a general problem here, since we do not want to assume
anything concerning the symmetries and/or the geometry of the
spacetime.

On the other hand, relativistic fluid dynamics has been studied 
in detail by
many authors (see e.g. \cite{CARTER,STEWART}), who usually do
not deal with a system of interacting fluids even if they
present a general formalism for a multi-constituent fluid. 
In fact, as will be explained later, the formalism
developped by Carter \cite{CARTER} for
multi-constituent perfect fluids does not overlap with the
description which we shall give here.

In this article, we stand in between the kinetic and the
fluid descriptions.
Note that even if such attempts have already been made,
they relied on a different approach to the problem. For instance,
Lindquist \cite{LINDQUIST} studied the
diffusion of photons
under the transport approximation (i.e. he studied
the evolution of particles flowing through an emitting
and absorbing medium described in macroscopic terms). 
Most of the studies of the transfer equation
are also based on moment methods \cite{CHERNIKOV}, 
on the ``grad's method of moment'' \cite{THORNE} or on
a spherical harmonic analysis \cite{ELLIS}. We will
not use such expansion methods here.

We will first use the microscopic approach (\& \ref{par2}) to relate
the interacting force and the source term to the collision term of the
Boltzmann equation (\& \ref{par3}).  In \& \ref{par5}, this result is
used to compute the force on a system of conducting fluids. We then
turn to the computation of the force for elastic collisions in
\& \ref{par6}, where we discuss the general form of the force and then
compute it for cases of cosmological interst (e.g.  photon-fermion
(using results on the photon-electron collision \cite{LARERE}) and
fermion-fermion scattering). \& \ref{par7} is devoted to the
computation of the general form of the source term and of the force
for inelastic collisions, namely fusion and fission, and we use it in
the case of photon (bremstrahlung) of fermion-antifermion annihilation
and of recombination (which is very important for a study of
decoupling in cosmology).

This general formalism is then used in to establish the equation of
evolution of a Compton scattering coupled photon-matter system in
Cosmology to linear order (\& \ref{par8}). All the theory of
multifluid linear cosmological perturbations \cite{KS} can be
recovered from our formalism which is non perturbative.

%------------------------------------------------------------------------ 
\section{Definition - notations - microscopic quantities} \label{par2}
%------------------------------------------------------------------------ 

The goal of this section is to introduce the distribution function and
its equation of evolution. For that purpose we need to introduce the
space on which such a function is defined. We finish by the
description of the 3+1 splitting with respect to an arbitrary vector
field.

\subsection{Distribution function} \label{spar21}

Let us first consider a single test particle with mass $m$ which moves in a 
gravitational field. Its motion is determined by the geodesic equation
\be
p^\mu=\frac{dx^\mu}{d\lambda};\quad \frac{Dp^\mu}{d\lambda}\equiv
\frac{dp^\mu}{d\lambda}+\Gamma^\mu_{\nu\rho}p^\nu p^\rho=0,
\label{MOTION}
\ee
where $\lambda$ is an affine parameter defined by the requirement that
$p^\mu$ be the 4-momentum. Hereafter, $\nabla_\mu$ denotes the
covariant derivative associated to the metric $g_{\mu\nu}$, whose
Christoffel symbols are $\Gamma^\mu_{\nu\rho}$.  Let us note that if
there are non gravitational forces (e.g. electromagnetic forces) then
we have to modify this equation (see \& \ref{par5}).

The rest mass of the particle is defined as
\be m^2=-p^\mu p_\mu.\label{MASS}\ee

Thus, according to (\ref{MOTION}), the state of the particle is determined
by the couple $(x^\mu,p^\mu)$ and the phase space is then
the tangent bundle over the spacetime manifold, i.e.
\be 
\T=\left\lbrace(\x,\p), \x\in{\cal M}, \p\in\T_x\right\rbrace,\label{TGBUN}
\ee
where ${\cal M}$ is the space-time and $\T_x$ is the tangent space
to ${\cal M}$ at $\x$. 
From now on, we will use the {\bf bold} style to denote quadri-vectors
when components notations are not needed (hence $\p\equiv p^\mu$
and $\p^2\equiv p^\mu p_\mu$ and greek indices run from 0 to 3).

The volume element on $\T_x$ supported
by the displacements $dp_1,dp_2,dp_3,dp_4$ (with components
$dp_1^\alpha$ etc.) is
\be 
\pi(\p)=\epsilon_{\alpha\beta\gamma\delta}
dp_1^\alpha dp_2^\beta dp_3^\gamma dp_4^\delta, \label{eVOL1}
\ee
 where $\epsilon_{\lambda\alpha\beta\gamma}$ is
the totally antisymmetric tensor such that $\epsilon_{0123}=\sqrt{-g}$.

We also define $\pi_+(\p)$, the volume element corresponding
to the subspace of $\T_x$ such that $p^\mu$ is non-spacelike 
and future directed,
\be 
\pi_+(\p)=H(-p_\mu u^\mu)H(-\p^2)\pi(\p),\label{EVOL11}
\ee
where $H$ is the heavyside function (i.e. $H(x)=0$ if $x<0$ and
$H(x)=1$ if $x>0$) and $u^\mu$ an arbitrary timelike vector field.

$\T_x$ is sliced in
hypersurfaces, $\P_m$, of constant $m$ called the mass-shell,
and defined by
\be 
\Pm(\x)=\left\lbrace\p\in\T_x, p^\mu p_\mu=-m^2,
p^\mu u_\mu>0\right\rbrace.\label{MASSHELL}
\ee
The volume element (\ref{eVOL1}) on $\T$ can then be decomposed on a
volume element, $m\pi_m$,  
on $\Pm$ by
\be
\pi_+(\p)=m\pi_m(\p)dm.\label{eVOL2}
\ee
The factor $m$ allows one to include particles of zero rest mass 
(see Ehlers \cite{EHLERS}). This defines the induced volume element
$m\pi_m(\p)$ on $\P_m$.

If we introduce an arbitrary future
directed unit timelike vector $u^\mu$ (i.e. satifying $u_\mu u^\mu=-1$),
the 3-volume supported by the three displacements $dx_1, dx_2, dx_3$ 
(with components $dx_1^\alpha$ etc.) in the hypersurface perpendicular
to $u^\mu$ is
\be 
dV(\u)=\epsilon_{\lambda\alpha\beta\gamma}u^\lambda
dx_1^\alpha dx_2^\beta dx_3^\gamma.\label{DVU}
\ee

We now consider a single fluid composed of particles which mass are a
priori different.  The distribution function, $f(\x,\p)$ will be
defined as the mean number of particles (on a statistical set) in a
volume $dV$ around $\x$ and $\pi(\p)$ around $\p$ measured by an
observer with 4-velocity $u^\mu$,
\be
dN(\x,\p)=f(\x,\p)(-p^\mu u_\mu)dV(\u)\pi(\p). \label{DEFF}
\ee
The assumptions involved in its existence have 
been discussed in details by Ehlers  \cite{EHLERS}.
Synge \cite{SYNGE} has demonstrated that $(-p^\mu u_\mu)dV(\u)$ is
independent of $u^\mu$, which implies that the distribution function is
a scalar. Moreover, $f(\x,\p)\geq0$ for all $x^\mu$ and all
allowed $p^\mu$. 

For a gas, $dN$ is the number of particles in a volume $dV\pi(\p)$ thus
the smoothness of $f$ depends on the existence of a
sufficient number of particles.

\subsection{Equation of evolution}\label{spar22}

The equations of motion (\ref{MOTION}) define on $\T$ an operator
called the Liouville operator (see e.g. \cite{DIU}) which reads
\be
{\cal L}=p^\mu\frac{\partial}{\partial x^\mu}+\frac{dp^\mu}{d\lambda}
\frac{\partial}{\partial p^\mu}=\frac{d}{d\lambda}, \label{DEFLIOU}
\ee
which characterises the rate of change of $f$ along the particle
worldlines. Using (\ref{MOTION}), this operator can be rewritten as
\be
{\cal L}[f]=p^\mu\partial_\mu f -\Gamma^\mu_{\nu\rho}
p^\nu p^\rho\frac{\partial}{\partial p^\mu}f.\label{DEFLIOU2}
\ee
The fact that the mass $m$ of the particle as defined in (\ref{MASS})
is a scalar function which is constant on each phase orbit leads to
\be
{\cal L}[m^2]=0.\label{MASSLIOU}
\ee

The Boltzmann equation states that this rate of change is equal to the
rate of change due to collisions, i.e. that
\be
{\cal L}[f]=C[f]. \label{DEFEQBOL}
\ee
$C[f]$ is the collision term and encodes the information about the
interactions between the particles of the fluid.\\

If we know consider a system of $N$ fluids (labelled by $i,j...$),
each of which is described by its distribution function $f_i(\x,\p)$, the
Boltzmann equation for a given fluid $i$ becomes
\be
{\cal L}[f_i]=\sum_{j}C_j[f_i,f_j]\equiv C_i[f_i],
\ee
$C_j[f_i,f_j]$ is the collision term  describing the
interaction between the fluid $i$ and the fluid $j$. For elastic
collisions, it must satisfy the symmetry
\be 
C_j[f_i,f_j]=C_i[f_i,f_j],\label{SIMCOLL}
\ee 
which means that in a collision between $i$ and $j$ the two distribution
functions undergo the same change. Following Israel and Stewart
\cite{IS2}, we will require that $C_j[f_i,f_j]$ is a local function
of the ``$f_i$" (i.e independent of their derivatives). 

\subsection{3+1 splitting}\label{spar23}

We perform a 3+1 splitting with respect to an arbitrary timelike
unit vector field $u^\mu$ (we discuss which vectors to use
according to the problem at hand in \& \ref{spar4}). The projection
tensor into the ``rest-space'' of an observer moving with this
4-velocity is defined by
\be
\perp_{\mu\nu}\equiv g_{\mu\nu}+u_\mu u_\nu.\label{PERP}
\ee
Any vector $p^\mu$ can be decomposed with respect to $u^\mu$ as
\be 
p^\mu=\lambda e^\mu+E u^\mu,\label{eSPLIT}
\ee
where the energy ($E$), the norm of the 
particles' 3-momentum ($\lambda$) and the direction with
respect to  $u^\mu$ are
\be 
E\equiv(-p^\mu u_\mu), \quad \lambda^2\equiv\perp_{\mu\nu}p^\mu p^\nu 
=E^2-m^2\quad{\rm and}\quad 
e_\mu\equiv\frac{\perp_{\mu\nu}p^\nu}{\sqrt{\perp_{\mu\nu}p^\mu p^\nu}},
\label{eLAMB}
\ee
so that $e^\mu$ and $u^\mu$ satisfy
\be 
u^\mu e_\mu=0,\quad e_\mu e^\mu=1.\label{PROPE}
\ee
We have dropped the $u$-dependence since there is no
ambiguity. However we will restore it if necessary (e.g
we will write $\perp^{\mu\nu}_u$). Here, we have split the 4-momentum $p^\mu$
but any other vector can be split according to the same procedure
(e.g. in equation \ref{SPLITVEC}).
Note that for a zero-mass particle these relations lead
to $\lambda=E$. \\

From these definitions and equations (\ref{EVOL11}) and (\ref{eVOL2}), 
we can reexpress the volume elements on
$\T_x$ and on the shell-mass  as (see \cite{EHLERS,ELLIS} for details)
\be \pi_+(\p)=m dm \lambda dE d\Omega \Longleftrightarrow
\pi_m(\p)=\lambda dE d\Omega \label{SPH},\ee
where $d\Omega$ is the solid angle spanned by three independent
$e^\mu$. The passage from (\ref{EVOL11}) to (\ref{SPH}) can be seen as the 
passage from a cartesian coordinate system to a spherical coordinate
system (i.e. we replaced the integrations on $dp^\alpha$ by an
integration on the norm and the angles.)\\

We also introduce the symmetric traceless tensor
\be \Delta_{\mu\nu}=e_\mu e_\nu-\frac{1}{3}\perp_{\mu\nu}\ee
which verifies
\be \Delta_{\mu\nu}\perp^{\mu\nu}=0,\quad
\Delta_{\mu\nu}g^{\mu\nu}=0,\quad\Delta_{\mu\nu}u^\mu=0\quad
{\rm and}\quad\Delta_{\mu\nu}e^\mu=\frac{2}{3}e_\nu.\ee
To finish, we give the following  useful integrals (see e.g. \cite{ELLIS}),
\be \int e^{\mu_1}...e^{\mu_n}\frac{d\Omega}{4\pi}=
\lb\begin{array}{ll}
0&n=2p+1\\
\frac{1}{n+1}\lp\perp^{(\mu_1\mu_2}...\perp^{\mu_{n-1}\mu_n)}\rp&
n=2p,\\ \end{array} \right.,
\ee
where $e^\mu$ satisfies equation (\ref{PROPE}). In particular we have
\bea 
\int \frac{d\Omega}{4\pi}&=& 1,\label{eINT1}\\
\int e^\nu e^\mu\frac{d\Omega}{4\pi}&=&\frac{1}{3}\perp^{\mu\nu},
\label{eINT2}\\
\int \Delta^{\alpha\beta}\Delta^{\gamma\delta}\frac{d\Omega}{4\pi}&=&
\frac{1}{45}\lp-2\perp^{\alpha\beta}\perp^{\gamma\delta}
+ 3\perp^{\alpha\gamma}\perp^{\beta\delta} + 3\perp^{\alpha\delta}
\perp^{\alpha\gamma}\rp.\label{eINT3}
\eea
We also choose the following conventions of symmetrisation
and anti-symmetrisation
$$A_{(\alpha_1...\alpha_n)}=\frac{1}{n!}\sum_{\sigma\in perm(1..n)}
A_{\sigma(1)...\sigma(n)}\quad{\rm and}\quad
A_{[\alpha_1...\alpha_n]}=\frac{1}{n!}\sum_{\sigma\in perm(1..n)}
\epsilon(\sigma)A_{\sigma(1)...\sigma(n)},$$
where $\epsilon(\sigma)$ is the signature of the permutation.

%------------------------------------------------------------------------ 
\section{Macroscopic quantities}\label{par3}
%------------------------------------------------------------------------ 

In the previous paragraph, we have given a microscopic description of
a set of interacting gases. The goal of this section is to
define a set of macroscopic quantities from the distribution function
and the collision term and then find the relations between these quantities.

\subsection{Definition}\label{spar1}

At any point $\x$, one can introduce, following
Ellis et al. \cite{ELLIS}, given a distribution function $f_i$ a set
of macroscopic quantities associated with each fluid $i$ by 
\be 
{X}_{i,a}^{\mu_1...\mu_n}(\x)=\int_{\T_x}\left(-p_\mu
p^\mu\right)^{a/2}p^{\mu_1}... 
p^{\mu_n}f_i(\x,\p)\pi_+(\p)=
\int_{m}\int_{\P_m}m^ap^{\mu_1}...
p^{\mu_n}f_i(\x,\p)m\pi_mdm \label{eDEF1}
\ee
and
\be 
{Y}_{i,a}^{\mu_1...\mu_n}(\x)=\int_{\T_x}\left(-p_\mu
p^\mu\right)^{a/2}p^{\mu_1}...
p^{\mu_n}C[f_i](\x,\p)\pi_+(\p)=
\int_{m}\int_{\P_m}m^ap^{\mu_1}...
p^{\mu_n}C[f_i](\x,\p)m\pi_mdm,\label{eDEF2}
\ee
where $m$ is the mass of the particles (defined in (\ref{MASS}))
and $a$ an integer. The particles of a given fluid can have different rest
mass (this is the case e.g. when one is dealing with a fluid of stars
or of galaxies).
We assume that each distribution function vanishes at infinity
on the mass shell quickly enough so that all these integrals
converge.

These quantities are clearly totally symmetric and follow (from equation 
(\ref{MASS})) the general recursion relation
\be  
g_{\mu\nu}X_{i,a}^{\mu_1...\mu_{n-2}\mu\nu}(\x)=
-X_{i,a+2}^{\mu_1...\mu_{n-2}}(\x).
\label{REC}
\ee
The functions $Y$ follow the same recursion relation.

The distribution function and the collision term are related
via the equation ({\ref{DEFEQBOL}) which implies  the two sets of quantities
$X$ and $Y$ are not independent.  
It can be shown (see Ehlers \cite{EHLERS}) by using the
Boltzmann equation (\ref{DEFEQBOL}), (\ref{MASSLIOU}) and that
$\partial_\mu \pi=\partial_\mu\lp\ln{\sqrt{-g}}\rp\pi(\p)=
\Gamma^\nu_{\mu\nu}\pi(\p)$ that the quantities $X$ and $Y$ are related by
\be 
\nabla_\mu X_{i,a}^{\mu_1...\mu_n\mu}=Y_{i,a}^{\mu_1...\mu_n},
\label{eDIV}
\ee
on which our following construction is based.

Among all these quantities, some are important in many applications,
\be
 n^\mu\equiv X_0^\mu,\quad N^\mu\equiv X_1^\mu\quad{\rm and}\quad
T^{\mu\nu}\equiv X_0^{\mu\nu}.\label{DEFN}
\ee
The vector $n^\mu$ is the number flux vector which is used
to define the average number flux velocity vector $v^\mu$ and the proper
density $n$ measured by an observer comoving with the fluid by
\be 
n^\mu=nv^\mu,\quad v^\mu v_\mu=-1.
\ee
Similarly, $N^\mu$ is the average mass flux vector which defines the mass
flux velocity vector $V^\mu$ and the energy
density ${\bar\rho}$ by \cite{EHLERS}
\be 
N^\mu={\bar\rho} V^\mu,\quad V^\mu V_\mu=-1.
\ee
For an observer with an arbitrary 4-velocity $u^\mu$, $n^\mu$ can be split
as
\be
n^\mu=nu^\mu+j^\mu;\quad n\equiv-(n_\mu u^\mu);\quad j^\mu=\perp^\mu_\nu j^\nu.
\label{SPLITVEC}
\ee
$j^\mu$ is the diffusion current as measured by this observer and vanishes
when $u^\mu=v^\mu$.

$T_{\mu\nu}$ is the energy-momentum tensor. It also determines a unique
average velocity, namely, its timelike eigenvector 
(see e.g. \cite{CHERNIKOV}). 
Whatever the timelike unit vector field $u^\mu$, chosen as
time direction, we can  
split the energy-momentum tensor under the general form
\be T_{\mu\nu}=\rho u_\mu u_\nu+P\perp_{\mu\nu}+ 2q_{(\mu}u_{\nu)}+
\pi_{\mu\nu},\label{eTMUNU}\ee
the quantities $\rho, P, q^\mu$ and $\pi_{\mu\nu}$ being defined as
\begin{equation}
\rho\equiv T_{\mu\nu}u^\mu u^\nu;\quad P\equiv\frac{1}{3}T_{\mu\nu}
\perp^{\mu\nu};\quad q_\mu\equiv-\perp^\nu_\mu T_{\nu\alpha} u^\alpha;\quad
\pi_{\mu\nu}\equiv \perp^\alpha_\mu\perp^\beta_\nu T_{\alpha\beta}-
P\perp_{\mu\nu}.\label{QPI}
\end{equation}
This decomposition is the most general splitting with respect to the
arbitrary vector field $u^\mu$ of a tensor of rank 2.
The two quantities $q_\mu$ and $\pi_{\mu\nu}$ are respectively called
the energy flux and the anisotropic stress and verify (from \ref{QPI})
\begin{equation} q_\mu u^\mu=\pi^\mu_\mu=\pi_{\mu\nu}u^\mu=0.
\end{equation}
By using the definition of $T_{\mu\nu}$ from the distribution function
(equation \ref{eDEF1}) and performing the splitting 
(\ref{eSPLIT}), a simple identification with equation (\ref{eTMUNU}) 
easily shows that
\bea
\rho(\x)&=&\int_0^\infty dm \int_m^\infty dE \int_\Omega f(\x,\p)E^2\lambda
d\Omega ,
\label{eRHO}\\
P(\x)&=& \frac{1}{3}\int_0^\infty dm \int_m^\infty dE\int_\Omega f(\x,\p)
\lambda^3 d\Omega , \label{eP}\\
q_\mu(\x)&=& \int_0^\infty dm \int_m^\infty dE\int_\Omega f(\x,\p) E\lambda^2 
e_\mu d\Omega\label{eQ}\\
\pi_{\mu\nu}(\x)&=& \int_0^\infty dm \int_m^\infty dE\int_\Omega f(\x,\p)
\lambda^3  \Delta_{\mu\nu}d\Omega.\label{ePI} 
\eea
All these quantities depend intrinsically on the choice of the vector
field $u^\mu$ via the splitting defined in equation
(\ref{eSPLIT}). Thus, $\rho$, $P$, $q^\mu$ and $\pi_{\mu\nu}$ will
respectively be the energy density, the pressure, the energy flux and
the anisotropic stress measured by an observer comoving with $u^\mu$.

\subsection{Macroscopic fluid dynamics}\label{spar4}

Using the equation (\ref{eDIV}) as well as the
definitions (\ref{eDEF1}-\ref{eDEF2}) and (\ref{DEFN}),
 we can relate the force
$F^\mu_i$ and the source term $\epsilon_i$ to the collision term defined
in (\ref{F3}-\ref{F4}) by
\be 
F^\mu_i(\x)\equiv Y_{i,0}^\mu=\int_{\T_x}p^\mu C_i[f_i](\x,\p)\pi_+(\p),
\label{eFORCE}
\ee
and
\be 
\epsilon_i(\x)\equiv Y_{i,0} =\int_{\T_x} C_i[f_i](\x,\p)\pi_+(\p).
\label{eEPS}
\ee

The Bianchi identies state that the total energy-monentum tensor is
conserved,which implies that we must have 
 the usual {\it action-reaction law}, i.e. that
\be 
\sum_{i} F_i^\mu=0,\label{eACT}
\ee
as long as there is no long range external force (such a force
can only be of electromagnetic origin \cite{CARTER}; this will be studied
in \& \ref{par5}).\\

Most of the time we will have to pick up a special frame to
compute the collision term. Some choices are possible even if
there are not compulsory.
\begin{itemize}
\item When a massless particle is interacting with a 
massive particle we will choose the rest frame of the massive particle
and thus
\be {\bar u}^\mu=\frac{p^\mu}{m}.\ee
\item When one has an elastic collision of two massive particles
we can use the center of mass rest frame defined by
\be P^\mu=p_1^\mu+p^\mu_2 \quad {\rm and}\quad
U^\mu=\frac{P^\mu}{\sqrt{-P^\mu P_\mu}}.\label{eDEFU}\ee
\end{itemize}
Besides these two velocities, we have seen that there exist some preferred
timelike vector fields associated with the motion of the matter
(e.g. $v^\mu$, $V^\mu$).

One could add to this dynamical description a thermodynamical description.
We have to emphasize here that we must give an equation of state to
close the system. This comes from the fact that when developped into moments,
at a given order the Boltzmann equation involves multipoles of higher
orders \cite{THORNE}. To close the system one has either
to troncate the system at a given order (with all the arbitrariness
it implies) or give an (or more if needed) equation of state for the fluid.
Even if we do not specify it, we assume that such an equation can be
given for concrete applications (see \& \ref{par8}).

The goal of the following section is to compute explicitely the
quantities (\ref{eFORCE}-\ref{eEPS}) in terms of the macroscopic
variables defined in section \ref{spar1}.

%------------------------------------------------------------------------- 
\section{Conducting fluid} \label{par5}
%------------------------------------------------------------------------- 

The easiest case where one can compute the force acting on a fluid
is the case of a conducting fluid in a electromagnetic field.
This has been computed for a single fluid by many authors (see \cite{STEWART}
for a review concerning electrodynamics in continuum media). 
We will just make the link between the 
microscopic and macroscopic approaches and show how useful the latter can be.
Let us start from the usual approach with a single fluid.

Since the total energy-momentum tensor, $T^{\mu\nu}_{em}$,
 is conserved, we have
\be 
\nabla_\nu T^{\mu\nu}=-\nabla_\nu T^{\mu\nu}_{em},
\ee
the electromagnetic field energy-momentum tensor being
defined by 
\be 
T^{\mu\nu}_{em}={\cal F}^{\mu\lambda}{\cal F}^\nu_\lambda
-\frac{1}{4}g^{\mu\nu}{\cal F}^2
,\ee
where ${\cal F}^{\mu\nu}$ is the electromagnetic tensor.
Using Maxwell's equations,
\be
\nabla_\nu{\cal F}^{\mu\nu}=j^\mu,
\ee
$j^\mu$ being the current density,
the electromagnetic force on the fluid is
\be 
F^\mu_B\equiv -\nabla_\nu T^{\mu\nu}_{em}={\cal F}^\mu_\alpha j^\alpha.
\ee

However in the case of a multi-fluid system, the only force that
can be computed with this method is the global
electromagnetic force on the system of fluids and not the force
on {\it each} fluid.\\

If we turn to the microscopic approach, then we have to
take into account the fact that, because of the Lorentz force,
the particles do not follow a geodesic between
two collisions. Their equation of motion is then given by
\be p^\mu_i\nabla_\mu p^\nu_i=e_i{\cal F}^{\mu\nu}p_\nu\quad
\Longleftrightarrow\quad\frac{dp^\mu_i}{d\tau}+\Gamma^\mu_{\nu\rho}
p^\nu_ip^\rho_i=e_i{\cal F}^{\mu\nu}p_\nu,\ee
where $e_i$ is the charge of the particles.

If we cast this relation in the Boltzmann equation as we did
in the section \ref{spar21}, it can be written
\bea
{\cal L}[f]=C_B[f]
\equiv-e_ip_\mu{\cal F}^{\mu\nu}\frac{\partial f}{\partial p^\nu}.
\eea
Hence the electromagnetic force acts as a collision term. In the
previous equation ${\cal L}$ stands for the Liouville
operator with no electromagnetic field, as defined in section \ref{spar21}. 

We can now compute the force coming from this collision term,
\bea
F_{B\rightarrow i}^\mu&=&-e_i\int p^\alpha p^\mu 
{\cal F}^\beta_\alpha \frac{\partial f}
{\partial p^\beta} \pi_+(\p), \nonumber \\
&=& e_i{\cal F}^\mu_\alpha n_i^\alpha \equiv {\cal F}^\mu_\alpha j_i^\alpha,
\eea
where we have performed an integration by part. Thus, the force acting
on a single fluid can be expressed in term of macroscopic quantities,
namely the electromagnetic tensor and the current density.

If we now compute the total electromagnetic force
on all the fluids we get
$$F^\mu_{B}=\sum_i F^\mu_{B\rightarrow i}={\cal F}^\mu_\alpha
\sum_i j^\alpha_i={\cal F}^\mu_\alpha j^\alpha,$$
which is the result we obtain with the macroscopic approach.

%-------------------------------------------------------------------------- 
\section{Elastic collisions}\label{par6}
%-------------------------------------------------------------------------- 

In this paragraph, we will compute the force $F^\mu$ for elastic
collisions. We begin by a derivation for binary collisions of
classical (in the sense that they are non quantum) particles, trying
to stay as general as possible. We then turn to the case of the
Compton scattering.  In the Thomson limit, even if electrons and
protons are not classical particles, the computation of the force
turns out to be an application of the general case.  However, in this
case, we can give the form of the corrections coming from the quantum
statistics.

We finish by a discussion on the range of validity of these
computations and determine the domain of applicability (range of
temperature) in which they can be used and then discuss the pertinence
of this formalism for cosmology.

\subsection{general case}\label{spar61}

We will give the most general form for the collision term for the process
$$ A(\p)+B(\q)\rightarrow A(\pp)+B(\qp).$$

Before computing the force, we will study its general symmetries,
which starts by the study of the kinematics of such a
collision. During the computation, we will have to break this symmetry
by working in the rest frame of one of the particles, and restore it
at the end.

We first work in the rest frame of
the center of mass and thus use the vector field $U^\mu$
defined in (\ref{eDEFU}) to
perform the splitting. We then have from equation (\ref{eLAMB})
\be 
E_A=\frac{m_A^2-p^\mu q_\mu}{P}\quad{\rm and}\quad
E_B=\frac{m_B^2-p^\mu q_\mu}{P}\quad{\rm with}\quad
P=(m_A^2+m_B^2-2p^\mu q_\mu)^{1/2},\label{eAB1}
\ee
\be
\lambda^2\equiv\lambda_A^2=\lambda_B^2=\frac{(p^\mu q_\mu)^2-m_A^2m_B^2}{P^2},
\quad {\rm and}\quad(-p^\mu q_\mu)=E_AE_B-\lambda_A\lambda_B.\label{eAB2}
\ee
$E_A$ and $E_B$ are the energies of the two particles $A$ and $B$ in
the center of mass rest frame, $\lambda$ is the amplitude of the
particles' 3-momentum is this frame.

The vectors $e_A^\mu$ and $e_B^\mu$, the direction of the 
ingoing particles, are given by
\bea e_A^\mu=-e_B^\mu&=&\frac{1}{\lambda P^2}\lb
(m_B^2-p^\nu q_\nu)p^\mu-(m_A^2-p^\nu q_\nu)q^\mu)\rb\nonumber\\
&=&\frac{1}{\lambda P}\lb E_Bp^\mu-E_Aq^\mu\rb.\eea

The general form for the collision term for binary collisions
of uncharged classical particle is \cite{TAUBER}
\be C[f_A](\p)=\int(-p^\nu q_\nu)\left[ f_A(\pp)f_B(\qp)-f_A(\p)f_B(\q)  
\right]
W(\p,\q,\pp,\qp)\pi_+(\q)\pi_+(\qp)\pi_+(\pp),\ee
where $W(\p,\q,\pp,\qp)$ is the probability of a collision 
$(\p,\q)\rightarrow(\pp,\qp)$.
General expressions for processes including electromagnetic effects and
quantum statistical effects have been proposed 
\cite{EHLERS,ISRAEL}. We will take such effects into account in the next
section (see equation (\ref{eQANT})). Here, we want to be as
general as possible and try not to describe the interaction
 in detail but give the general form of the force.
The microscopic reversibility of the collision imposes the 
symmetry
\be W(\p,\q,\pp,\qp)=W(\pp,\qp,\p,\q).\ee

Before we estimate $W(\p,\q,\pp,\qp)$, let us study the symmetry of the
force
\be F^\mu_{B\rightarrow A}=
\int(-p^\nu q_\nu)\lb f_A(\pp)f_B(\qp)-f_A(\p)f_B(\q) \rb p^\mu
W(\p,\q,\pp,\qp)\pi_+(\q)\pi_+(\qp)\pi_+(\pp)\pi_+(\p).\ee If we sum
the two forces (i.e. $F^\mu_{B\rightarrow A}$ and $F^\mu_{A\rightarrow
B}$) and set $ W(\p,\q,\pp,\qp)=\delta^{(4)}({\bf P}-{\bf P'})
R(\p,\q,\pp,\qp)$, where $R$ and $W$ have the same symmetry, in order
to make the conservation of energy-momentum explicit ,then since
$P^2=P^{\prime2}$ implies that $(-\p\q)=(-\pp\qp)$, the integrand of
$F^\mu_{B\rightarrow A}+F^\mu_{A\rightarrow B}$ will be antisymmetric
in the transformation $(\p,\q)\rightarrow(\pp,\qp)$ and thus
\be F^\mu_{B\rightarrow A}+F^\mu_{A\rightarrow B}= 0.\label{eSYM}\ee  
This was expected by construction from (\ref{eACT}) but had to be checked on
the general form.  \\

To compute the force, we must relate the $W(\p,\q,\pp,\qp)$ probability 
to the differential cross section $d\sigma^{pq\rightarrow p'q'}$ defined as
\be W(\p,\q,\pp,\qp)\pi_+(\qp)\pi_+(\pp)=(-p^\mu q_\mu)
d\sigma^{pq\rightarrow p'q'}.\ee
The cross section can be decomposed \cite{IZ} as the product of
a matrix element $M^{kp\rightarrow k'p'}$ and of a two body phase space 
element $\lp D^{kp\rightarrow k'p'}\rp$ as
\be
d\sigma^{pq\rightarrow p'q'}=\frac{1}{4}
\frac{1}{(-p^\mu q_\mu)}|M^{kp\rightarrow k'p'}|^2 
\lp{D^{kp\rightarrow k'p'}}\rp^2,\ee
with
\be \lp{D^{kp\rightarrow k'p'}}\rp^2=(2\pi)^4\delta^{(4)}(\p+\q-\pp-\qp)
\pi_+(\qp)\pi_+(\pp).\ee

If we consider massive particles with respective rest mass
$m_A$ and $m_B$, one can convince oneself that the matrix element
has to be of the form
\be |M^{kp\rightarrow k'p'}|^2=\sigma\Upsilon(m_A^2,m_B^2,p^\mu q_\mu,
e^\mu,e^{\prime\mu}),\ee
where $\sigma$ is the scalar cross section and the function 
$\Upsilon$ can be decomposed as
\be\Upsilon=\alpha+\beta e^\mu e'_\mu +\gamma\Delta^{\mu\nu}
\Delta_{\mu\nu}'+...,\label{eTAYLOR}\ee
where the coeficients $\alpha, \beta, \gamma$ depend on $m_A^2$, $m_B^2$ and
$p_\nu q^\mu$ and
where $e^\mu$ and $e^{\prime\mu}$ are the directions of the 
ingoing and outgoing particles. This form comes from a multipole
expansion of the matrix element in which the coefficients
depend only on the scalar invariants of the collision
(an example of such a function $\Upsilon$ is given
in equation (\ref{eUPSI})). As  will be shown below,
these are the only relevant terms
to compute the force since higher multipoles will not contribute.

We will now assume that the particles are either non-relativistic
in the center of mass rest frame or that one of the two particles is
massless. The first approximation can be stated by
assuming that the 3-momentum ($\lambda$ defined in (\ref{eAB2})) is small 
compared to the total energy ($E_A$ and/or $E_B$) which is of same
order of magnitude that the rest mass ($m_A$ and/or $m_B$), i.e.
that
$$\lambda\ll E\sim m.$$
Since $\Upsilon(m_A^2,m_B^2,p^\mu q_\mu)=\Upsilon(m_A^2,m_B^2,\lambda,
E_A,E_B)$,
in this approximation we can Taylor expand each coefficient of
equation (\ref{eTAYLOR}) in power of $\lambda/m_{AB}$, 
with $m_{AB}=inf(m_A,m_B)$
as (on the example of $\alpha$)
$$\alpha(m_A^2,m_B^2,p^\mu q_\mu)=\alpha_{NR}+\frac{\lambda}{m_{AB}}{\tilde
\alpha}+O\lp\lp\frac{\lambda}{m_{AB}}\rp^2\rp.$$
On the other hand if we assume that one of the particle
is ultra-relativistic, then we can make a ``Thomson-like"
approximation, which says that the energy of the zero-mass particle
is small compared with the rest mass of the particle it
scatters with (i.e. $p^\mu q_\mu\ll m^2_{AB}$ with $m_{AB}=sup(m_A,m_B)$).
Each coefficient of equation (\ref{eTAYLOR}) can then be expanded as
$$\alpha(m_A^2,m_B^2,p^\mu q_\mu)=\alpha_{UR}
-\frac{p^\mu q_\mu}{m^2_{AB}}{\hat
\alpha}+O\lp\lp\frac{p^\mu q_\mu}{m^2_{AB}}\rp^2\rp.$$
Technically these two approximations reduce the knowledge
of the function $\Upsilon$ to a set of scalars (
$(\alpha_{NR},\beta_{NR},\gamma_{NR},
{\tilde\alpha},{\tilde\beta},{\tilde\gamma})$
in the first approximation and
$(\alpha_{UR},\beta_{UR},\gamma_{UR},{\hat\alpha},{\hat\beta},{\hat\gamma})$
in the second one).

Since at zeroth order the two approximations lead to the
same form of the force, we will first compute this term and then evaluate
the first order correction in the two approximations. Note that even 
if the form of the force is the same at zeroth order, the force
itself will be different because $\alpha_{NR}$ and $\alpha_{UR}$
are not the same constants. We have
$$\alpha_{NR}=\alpha(m_A^2,m_B^2,E_A=m_A^2,E_B=m_B^2,\lambda=0),$$
$$\alpha_{UR}=\alpha(m_A^2,m_B^2,p^\mu q_\mu=0).$$
For the sake of simplicity we set $\alpha=(\alpha_{NR},\alpha_{UR})$.\\

We will compute $F^\mu_{AB}$ in the rest frame of $B$. 
As explained at the beginning of this section, this 
breaks the symmetry between $A$ and $B$ (see (\ref{eSYM})). Therefore,
we will have to restore this ``hidden" or ``lost" symmetry
at the end of the computation. Another solution would have been
to split the force into two halves and compute one part in the
rest frame of $A$ and the other half in the rest frame of $B$.
Let us note that this symmetry can be restored only in the first
approximation where we can compute the force either
in the rest frame of $A$ or the rest frame of $B$. In the
second approximation, this is no longer possible and the
force will not be obviously symmetric.
We define $F^\mu_{AB}$ by
\be F^\mu_{AB}=
\int\left[(-p^\nu q_\nu)\lb f_A(\pp)-f_A(\p)\rb f_B(\q) p^\mu
\Upsilon(m_A^2,m_B^2,p^\mu q_\mu,e^\mu,e^{\prime\mu}) \frac{d\Omega'}{4\pi}
\pi_+(\q)\lambda dEd\Omega\right]_{rfB},
\label{eFORCE2}\ee
where ``$rfB$" means that the quantities are evaluated in the
rest frame of $B$. From equation (\ref{eSYM}), it follows that 
in the first approximation (i.e. the non-relativistic approximation) we have
$$F^\mu_{A\rightarrow B}= F^\mu_{[AB]},$$
(where $[AB]$ means that we anti-symmetrise on $A$ and $B$ the expression
(\ref{eFORCE2}))
and in the second approximation (the ultra-relativistic approximation)
$$F^\mu_{A\rightarrow B}= F^\mu_{AB},\quad F^\mu_{B\rightarrow A}
= -F^\mu_{AB}.$$

We perform the splitting (\ref{eSPLIT}) of $p^\mu$ with
respect to $u^\mu$ defined by $u^\mu=\frac{q^\mu}{m_B}$,
and inject in the integral (\ref{eFORCE2}). For the sake of clarity
we will split $F^\mu_{AB}$ in $F^{(0)\mu}_{AB}+F^{(1)\mu}_{AB}$
where $(0)$ and $(1)$ refer to the zeroth and first orders
in the expansion of $\Upsilon$ either in ``$\lambda/m_{AB}$" or
in ``$(-pq)/m_{AB}^2$".

Thus, taking into account the fact that $q^\mu e_\mu=0$ and
that $\int e^\mu d\Omega=\int \D^{\mu\nu}d\Omega=0$, it follows
\bea
F^{(0)\mu}_{AB}&=&\sigma\alpha\int p^\mu p^\nu q_\nu f_A(\p)f_B(\q)
\pi_+(\q)\pi_+(\p)\nonumber\\
&+&\sigma\int\lp E^2q^\mu+E\lambda m_Be^\mu\rp f_B(\q) \pi_+(\q)\nonumber\\
&&\lb\alpha\int f_A(\pp)\frac{d\Omega'}{4\pi}+\beta e^\alpha
\int f_A(\pp)e'_\alpha\frac{d\Omega'}{4\pi}+\gamma\Delta^{\alpha\beta}
\int f_A(\pp)\Delta_{\alpha\beta}'\frac{d\Omega'}{4\pi}\rb
\lambda dEd\Omega.
\eea
Using the definitions (\ref{eRHO}-\ref{ePI}) and the integrals 
(\ref{eINT1}-\ref{eINT3}), this
reduces to
\be
F^{(0)\mu}_{AB}=\sigma\lb \alpha n_{B\nu}T^{\mu\nu}_A+\alpha\rho_A n^\mu_B+
\frac{\beta}{3m_B}T_{B\nu}^\nu q^\mu_A \rb.\label{eF0}\ee
Let us recall that $q^\mu_A$ is the energy flux with respect to $n^\mu_B$, 
i.e with respect to the unit vector $u^\mu_B$ colinear to
$n^\mu_B$,
$$q^\mu_A=-\lb T^{\mu\nu}_A u_{B\nu}+T_A^{\alpha\beta}
u_{B\alpha}u_{B\beta} u^\mu_B\rb.$$
Thus, we have obtained the expression of the force on the fluid composed
of particles $A$ in terms of the macroscopic quantities describing the
two fluids and of the coefficients ($\alpha,\beta,\gamma,\sigma$) describing
the collision.\\

Let us now turn to the evaluation of the first order
corrections. We begin by the ``non-relativistic"
approximation,
\bea
F^{(1)\mu}_{AB}&=&\frac{\sigma}{m_{AB}}{\tilde\alpha}\int p^\mu p^\nu 
q_\nu f_A(\p)f_B(\q)\lambda\pi_+(\q)\pi_+(\p)\nonumber
+\frac{\sigma}{m_{AB}}\int\lp E^2q^\mu+E\lambda m_Be^\mu\rp 
f_B(\q) \pi_+(\q)\nonumber\\ 
&&\lb
{\tilde\alpha}\int f_A(\pp)\frac{d\Omega'}{4\pi}+{\tilde\beta} e^\alpha 
\int f_A(\pp)e'_\alpha\frac{d\Omega'}{4\pi}+{\tilde\gamma}\Delta^{\alpha\beta}
\int f_A(\pp)\Delta_{\alpha\beta}'\frac{d\Omega'}{4\pi}\rb
\lambda^2 dEd\Omega.
\eea
Using the definition of $\lambda$ (\ref{eLAMB}), and the set
of new quantities defined by
\bea
J_A&=&\int_m^\infty\int_\Omega f_A(\x,\p)\lambda^2 E^2dEd\Omega,\\
J_A^\alpha&=&\int_m^\infty\int_\Omega f_A(\x,\p)\lambda^3 Ee^\alpha 
dEd\Omega,\label{eJA}\\
J_A^{\alpha\beta}&=&\int p^\alpha p^\beta p^\gamma e_\gamma
f_A(\x,\p)\pi_+(\p),
\eea
and the definitions (\ref{eRHO}-\ref{ePI}) and integrals 
(\ref{eINT1}-\ref{eINT2}),
we obtain
\be
F^{(1)\mu}_{AB}=\frac{\sigma}{m_{AB}}\lb{\tilde\alpha}n_B^\nu J_A^{\mu\nu}
+{\tilde\alpha} J_An_B^\mu+{\tilde\beta} T_{B\nu}^\nu J_A^\mu\rp.
\label{eF1A}\ee

We will now turn to the second approximation (i.e. 
$p^\mu q_\mu\ll m^2_{AB}$) and evaluate 
the first order contribution. It reads
\bea
F^{(1)\mu}_{AB}&=&\frac{\hat\sigma}{m_{AB}^2}\alpha\int p^\mu p^\nu 
p^\lambda q_\nu q_\lambda f_A(\p)f_B(\q)\pi_+(\q)\pi_+(\p)
+\frac{\sigma}{m_{AB}^2}\int\lp m_BE^3q^\mu+E^2\lambda m^2_Be^\mu\rp 
f_B(\q) \pi_+(\q)\nonumber\\ 
&\lb\right.&\left.
{\hat\alpha}\int f_A(\pp)\frac{d\Omega'}{4\pi}+{\hat\beta} e^\alpha 
\int f_A(\pp)e'_\alpha\frac{d\Omega'}{4\pi}+{\hat\gamma}\Delta^{\alpha\beta}
\int f_A(\pp)\Delta_{\alpha\beta}'\frac{d\Omega'}{4\pi}\rb
\lambda dEd\Omega,
\eea
and thus
\be
F^{(1)\mu}_{AB}=\frac{\sigma}{m_Am_B}\lb{\tilde\alpha}N_B^\mu I_A
+\frac{\tilde\beta}{3}\perp^{\mu\alpha}_BT_{B\mu}^\mu I_{A\alpha}
-{\tilde\alpha}T_{B\nu\lambda}X_{0A}^{\mu\nu\lambda}\rb,
\label{eF1B}\ee
where we have introduced the two macroscopic quantities
\bea
I_A=\int_m^\infty\int_\Omega f_A\lambda E^3dEd\Omega,\quad
I_A^\alpha=\int_m^\infty\int_\Omega f_A\lambda^2 E^2e^\alpha dEd\Omega,
\eea
and where the indice $B$ in $\perp^{\mu\alpha}_B$ means that we used the
4-velocity $u_B^\mu$. $X_{0A}^{\mu\nu\lambda}$ is defined in (\ref{eDEF1}).

In conclusion the force is given by
\bea
F^\mu_{B\rightarrow A}&=&2\lp F^{(0)\mu}_{[AB]}+F^{(1)\mu}_{[AB]}\rp
\nonumber\\
F^\mu_{B\rightarrow A}&=&\lp F^{(0)\mu}_{AB}+F^{(1)\mu}_{AB}\rp  
,\eea
according to the first or the second approximation. $F^{(0)}$
is given by (\ref{eF0}) an $F^{(1)}$ either by (\ref{eF1A}) or
(\ref{eF1B}).

Let us stress that we could go on in the expansion of $\Upsilon$ and compute
corrections to the force at different orders. This would however
involve the introduction of new macroscopic quantities (like e.g. 
$J_A$, $J_A^\alpha$...) and of tensors of higher rank 
(like e.g. $X_{0A}^{\mu\nu\lambda}$...). This is an example 
of what we mentioned in \& \ref{spar4} 
since $F^{(0)}$ involves $q^\mu$ and $F^{(1)}$, $X_0^{\lambda\mu\nu}$.

In the case of quantum particles, we have to take into account
quantum statistics effects which will be evaluated in the
next section.

\subsection{Photon-electron scattering} \label{spar62}

We will focus here on the elastic Compton scattering between
electrons and photons,
$$e^-(\p)+\gamma(\k)\rightarrow e^-(\pp)+\gamma(\kp).$$
We will try to follow the general computation that we have
developped in the former paragraph. However we have to take
into account quantum statistics.
The general form of the collision term (see e.g. 
\cite{MASAKI,OXENIUS,MIHALAS}) is
\bea C[f_\gamma](\x,\k)&=&\int (-p^\mu k_\mu) \pi_+(\p)
\lb f_e(\pp)f_\gamma(\kp)\lp 1+\frac{f_\gamma(\k)}{2}\rp
\lp 1-\frac{f_e(\p)}{2}\rp\right.\nonumber\\ 
&&-
\left. f_e(\p)f_\gamma(\k)\lp 1+\frac{f_\gamma(\kp)}{2}\rp
\lp 1-\frac{f_e(\pp)}{2}\rp \rb
d\sigma^{kp\rightarrow k'p'},\eea
where the factors 
$\lp 1+\frac{f_\gamma(\k)}{2}\rp$ and $\lp 1-\frac{f_e(\p)}{2}\rp$ are
terms coming from the Bose-Einstein and Pauli statistics for
the photon and electron respectively\cite{DIU}. The differential cross section
for the Compton scattering is given by\cite{IZ}
\be
d\sigma^{kp\rightarrow k'p'}=\frac{1}{4}
\frac{1}{(-p^\mu k_\mu)}|M^{kp\rightarrow k'p'}|^2 
\lp{D^{kp\rightarrow k'p'}}\rp^2,\ee
where $|M^{kp\rightarrow k'p'}|$ and $\lp{D^{kp\rightarrow k'p'}}\rp^2$ 
are respectively the matrix element and the two bodies phase space element.
If we now work in the reference frame of the electron, we have to
choose
\be u^\mu=\frac{p^\mu}{m_e},\ee
to perform the splitting (\ref{eSPLIT}),
from which it follows that the coefficients $|M^{kp\rightarrow k'p'}|^2$
and $\lp{D^{kp\rightarrow k'p'}}\rp^2$  are respectively given by
\be \lp{D^{kp\rightarrow k'p'}}\rp^2
=(2\pi)^4\delta^{(4)}(\p+\k-\pp-\kp)\pi_+(\pp)\pi_+(\kp)
=\frac{1}{4}(2\pi)^{-2}\frac{E}{m_e}\lp\frac{E'}{E}\rp^2d\Omega',\ee
and
\be
|M^{kp\rightarrow k'p'}|^2=16(2\pi)^2m_e^2
\lp\frac{E}{E'}\rp^2\lb1+\Delta_{\mu\nu}\Delta^{\mu\nu\prime}
+\frac{3}{4}\lp\frac{E}{E'}+\frac{E'}{E}-2\rp\rb\frac{\sigma_T}{4\pi}.
\label{eUPSI}\ee
$\sigma_T$ is the Thomson scattering cross section. The differential cross 
section is then 
\be d\sigma^{kp\rightarrow k'p'}=\lb1+\frac{3}{4}
\Delta_{\mu\nu}\Delta^{\mu\nu\prime}   
+\frac{3}{4}\lp\frac{E}{E'}+\frac{E'}{E}-2\rp\rb\frac{\sigma_T}{4\pi}
d\Omega'.\label{eDIFF}\ee
If we use the conjugate
process (i.e $kp'\rightarrow k'p$), we can factorise $f_e(\p)$.

We will also make two following approximations
\begin{itemize}
\item we will take the Thomson limit of the Compton
scattering, which implies that
\be \frac{E}{m_e}\sim\frac{E'}{m_e}\ll1,\ee
\item and we will neglect the quantum statistics.
\end{itemize}

With these approximations the collision term can be reset as
\be 
C_T[f_\gamma](\x,\k)=\sigma_T\int (-p^\mu k_\mu) \pi_+(\p)      
f_e(\p)\lp 1+\frac{3}{4}\Delta_{\mu\nu}\Delta^{\mu\nu\prime} \rp
\lp f_\gamma(\kp)-f_\gamma(\k)\rp\frac{d\Omega'}{4\pi},\ee
from which it follows that the force is
\be F^\nu(x^\mu)=\int C_T[f_\gamma](\x,\k) k^\nu \pi_+(\k).\ee
We can compute this force by using the integrals (\ref{eINT1}-\ref{eINT3}) 
given in 
\& \ref{par2} and the macroscopic quantities (\ref{eRHO}-\ref{ePI}), 
we obtain
\be F_{e\rightarrow\gamma}^\nu
=\sigma_T\lp n_{e\mu} T^{\mu\nu}_\gamma+\rho_\gamma
n_e^\nu\rp.\ee
Note that this result could have been obtained from
the general derivation of the section \ref{spar61} in
the second approximation (i.e  by making $m_B=m_e\rightarrow+\infty$
in (\ref{eF0})). The force on
the electron is given by
\be F_{\gamma\rightarrow e}^\nu=-F_{e\rightarrow\gamma}^\nu.\ee

We now need to compute the corrections to this force. They
are of two origins, the corrections coming from the
fact that we do not take the Thomson limit (i.e. corrections
in $E/m_e$ and $E'/m_e$) and corrections coming
from quantum effects (i.e. terms in $ff$). The first one
are of the same kind that the one for
the diffusion (section \ref{spar61}). Equation (\ref{eDIFF}) tells us 
that (${\hat\beta}=0,{\hat\gamma}=0,{\hat\alpha}=3/2$) and thus,
according to (\ref{eF1A}), the correction is given by
\be
F^{(1)\mu}_{e\rightarrow\gamma}=\frac{3}{2}\frac{\sigma_T}{m_e^2}
\lb N_e^\mu I_\gamma
-T_{e\nu\lambda}S_\gamma^{\mu\nu\lambda}\rb\label{eFEP}\ee
where we have introduced the macroscopic quantity
\be
I_\gamma=\int_m^\infty\int_\Omega f_AE^4dEd\Omega,
\ee
We will now deal with the quantum statistics corrections in the
Thomson limit. Using the general expression of the collision
term we have
\bea
F^{(quant)\mu}_{e\rightarrow\gamma}&=&\frac{\sigma_T}{2}\int(-p^\nu k_\nu)
\lb f_e(\p)\lp f_e(\pp)f_\gamma(\k)-f_\gamma(\k)f_\gamma(\k)-f_e(\pp)
f_\gamma(\kp)\rp+f_e(\pp)f_\gamma(\kp)f_\gamma(\k)\rb \nonumber \\
&&\lp1+\frac{3}{4}\D^{\mu\nu}\D'_{\mu\nu}\rp  k^\mu 
\delta^{(4)}(\p+\k-\pp-\kp)\pi_+(\p)\pi_+(\pp)\pi_+(\k)\pi_+(\kp). 
\label{eQANT}
\eea
To compute this integral, we need to define a whole set of macroscopic
quantities related to the moments of ``$f^2$". We will not
go further here, but we see that such a computation is possible.

\subsection{Photon-baryon scattering} \label{spar63}

If we still work in the Thomson limit, which is equivalent to assume that
the electron mass is infinite, the collision
term coming from the scattering of photons by baryons will
be the same and thus
\be F_{p\rightarrow\gamma}^\nu=\sigma_T\lp n_{p\mu} 
T^{\mu\nu}_\gamma+\rho_\gamma n_p^\nu\rp\label{eFMPHOT}.\ee

If we have other fermions in the problem, all
interaction between photons and these fermions will be
described by the same force since we
stay in the Thomson limit. The first order term will
differ from one fermion to the other since it is
proportional to $m_f^2$ as seen on (\ref{eFEP}).

\subsection{Discussion of the domain of validity}\label{spar64}

Three approximations have to be discussed, namely
\begin{itemize}
\item $(-p^\mu q_\mu)\ll m_{AB}^2$ or $\lambda\ll m_{AB}$, i.e 
the fact that the particles are treated either as non relativistic or
ultra relativistic,
\item the classical approximation, i.e. the fact that we have
neglected the quantum factors coming from Bose-Einstein and
Fermi-Dirac  statistics,
\item the gas approximation
\end{itemize}

We have assumed that the particles were non relativistic in the rest
frame of the center of mass, which ammounts to assuming that the
temperature was not too high since we described the constituents by
massive particles.  For a given element this gives a maximum
temperature $$\Theta_i<\Theta_{i*}=\frac{m_ic^2}{k_B},$$ $k_B$ being
the Boltzmann constant.  The index $i$ emphasizes the fact that the
fluids do not have to be in thermal equilibrium and can have different
temperature (moreover this temperature is defined as the statistical
temperature).  For electrons, we have $\Theta_{e*}=6.10^9K$.  However,
this limit can be relaxed if we take into account the corrections in
``$\lambda/m$". For higher temperature, electrons can be treated like
a radiation fluid, which means that we then used the
``ultra-relativistic'' approximation.

If we turn to quantum effects, the temperature must not be too low, in
order that the medium be non degenerate. For fermions of rest mass $m$
and of spin $s$, this condition (see e.g. \cite{DIU}) is satisfied if
$$T>\frac{2\pi\hbar^2}{k_Bm}\lp\frac{n}{2s+1}\rp^{2/3}.$$ In a
Friedmann-Lema\^ \i tre universe with $\Omega_0=1$, the density is
equal to the critical density $\rho_{c0}\sim 10^{-29}h^2g.cm^{-3}$,
where $h$ is related to the Hubble constant via $H_0=100 h
km.s^{-1}.Mpc^{-1}$. This leads to an average particles density of
$n_{0}\sim 1~particle.cm^{-3}$ for the matter.  Thus, for electrons
($s=1/2$, $m=m_e$) the former condition and the assumption that
$n_{e0}\sim n_0$ gives
$$\lp\frac{T}{T_0}\rp>10^{-16}\lp\frac{\rho_e}{\rho_{e0}}\rp^{2/3},$$
where $T_0\equiv 2.7K$ and where we have assumed $h\sim1$.  Assuming
that the electrons behave like matter, so that they scale like $a^3$,
$a$ being the scale factor of the universe, this can be rewritten in
term of redshift as $$z<10^{16}.$$ For this range of redshifts the gas
of photons (or electrons) will behave like a classical perfect gas. We
see that we have to treat electrons like radiation before we have to
take into account the quantum effects.  The limit for bosons is of the
same order of magnitude.

Since we have computed the general force between uncharged classical
particles (section \ref{spar61}), the formalism can be applied, for
instance, to fluid of galaxies in a cluster or of stars in galaxies
etc.. In such a case one cannot assume that the fluid is composed of
particles with the same rest mass and each fluid will have a mass
spectrum. Since we perform all the integrations on ${\T}_x$ instead of
the mass shell, all the previous results still stand. Note that the
approximation of non relativistic relative speed is a very good
approximation for this kind of fluids.

The last point we need to discuss concerns the gas approximation,
i.e. of the splitting of the total fluid (i.e. of energy-momentum
tensor) into individual fluids. For that purpose we will compare the
mean free path, $l_c$ say, and the average distance between two
particles, $d$ say, for the electrons. On the one hand,
$$l_c\sim\frac{1}{n\sigma_T}\sim\frac{1.5\times10^{24}cm}{(1+z)^3},$$ where
$\sigma_T$ is the Thomson scattering cross section ($\sigma_T\sim
6.65\times10^{-25} cm^{-2}$). On the other hand, since 
$d\sim n_e^{-1/3}$, using
the previous approximate value of $n_{e0}$, we have $$d\sim
\frac{1cm}{1+z}.$$ The gas approximation is valid if
$$l_c>d\Longleftrightarrow z<1.22\times10^{12}.$$ Thus, from the time of
electron-positron annihilation ($T\sim10^{10}K$,i.e. $z\sim3\times10^9$) until
the time of recombination of hydrogen, it is a very good approximation
\cite{WEINBERG} to treat the content of the universe as a
nonrelativistic gas plus blackbody electromagnetic radiation.

Moreover, we know from the theory of the strong interaction that
thanks to the ``asymptotic freedom", the concept of
weakly interacting particles is appropriate for very dense
systems \cite{IZ}. Thus, beyond its conventional
range of applicability discussed just above, the hypothesis
of weakly interacting particles and thus their description
by the kinetic theory may be extended to early universe.

As we see on this discussion, it is a crucial point of the theory
to assume that kinetic theory can be applied and 
this will have to be checked on any particular case.

%-------------------------------------------------------------------------- 
\section{Inelastic collisions} \label{par7}
%-------------------------------------------------------------------------- 

Source terms can arise from many phenomena. For instance if there
are some unstable particles in the problem, we must take into account
decay, fission and fusion. In the early universe matter
and anti-matter coexisted, which drives us to consider
fermion-anti-fermion annihilation.
When one turns to photons, one
major effect has to be considered, namely the Bremstrahlung
and the recombination. We finish by relating our
approach to other work on particles production.

\subsection{General case of fission}\label{spar71}

We will try to give the most general term of the source term for
fission
$$A(\p)\rightarrow B(\pp)+C(\qp).$$

As for the scattering, we start by the kinematics. We work in the rest frame
of the decaying particle and thus
\be u^\mu=\frac{p^\mu}{m_A}.\ee
It is obvious to see that $E_A=m_A^2$ and $\lambda_A=0$. This
implies that the vector $e^\mu\equiv e_A^\mu$ is an 
arbitrary unit vector. The vectors $p^{\prime\mu}$ and $q^{\prime\mu}$
are then given by
\bea
p^{\prime\mu}&=&\frac{1}{2}\lp 1+\delta_{ABC}\rp p^\mu+\frac{1}{2}
\lp 1-\alpha_{ABC}+\delta_{ABC}\rp^{1/2}e^\mu \label{ePPRIM}\\
q^{\prime\mu}&=&\frac{1}{2}\lp1-\delta_{ABC}\rp p^\mu-\frac{1}{2}
\lp1-\alpha_{ABC}+\delta_{ABC}\rp^{1/2}e^\mu,\eea
with $\delta_{ABC}\equiv(m_B^2-m_C^2)/m_A^2$ and $\alpha_{ABC}
\equiv(m_B^2+m_C^2)/m_A^2$. The fission is possible if
and only if $m_A\geq m_B+m_C$.

In fact fission is the easiest case of non elastic
collision. Since $p^\mu$ is the only vector of the
problem and the emission is isotropic in the rest frame
of $A$, it is obvious that
\be C[f_A]=C(-\p^2)=-\tau_A^{-1},\ee
where $\tau_A$ is a constant  representing the lifetime of A.
Thus, $C[f_A]$ represents the probability of
decay of A per unit time and the Boltzmann equation with such
a collision term (${\cal L}[f]=-f/\tau$) describes the relaxation
toward the equilibrium solution $f_{eq}=0$.

The source term is then given by
\be\epsilon_A=\int C[f_A]\pi_+(\p)=-\frac{n_A}{\tau_A}.\ee
The production of $B$ and $C$ are related to $\epsilon_A$ by
\be \epsilon_B=\epsilon_C=-\epsilon_A.\ee
The force is 
\be F^\mu_{\rightarrow A}=\int C[f_a]p^\mu\pi_+(\p)
=-\frac{n^\mu_A}{\tau_A}.\ee  
It can be understood as a ``rubbing'' coming from
the decay.The force on $B$ and $C$ is given by
\be F^\mu_{\rightarrow B}=\int p^{\prime\mu}\frac{f_A(\p)}{\tau_A}
\pi_+(\pp),\ee
where $\pp$ and $\p$ are related by the kinematic relations. If we
cast (\ref{ePPRIM})
in the equation of $F^\mu_{\rightarrow B}$
and use the fact that the emission is isotropic in the rest frame
of $A$ (since $e^\mu$ is arbitray), we get that
\be F^\mu_{\rightarrow B}=\frac{1}{2}\lp 1+\delta_{ABC}\rp
n^\nu_A/\tau_A\quad{\rm and}\quad 
F^\mu_{\rightarrow C}=\frac{1}{2}\lp 1-\delta_{ABC}\rp 
n^\nu_A/\tau_A.\ee
It can be checked that, as expected,
$$ F^\mu_{\rightarrow A}+F^\mu_{\rightarrow B}+F^\mu_{\rightarrow C}=0.$$

We must stress that this source term and these forces were computed
without any approximation.

\subsection{General case of fusion}\label{spar72}

We will try to give the most general term of the source term for
fusion
$$A(\p)+B(\q)\rightarrow C(\qp).$$

The kinematics of the fusion is analogous to the one of
the fission. If we work in the
rest frame of particle $C$, we have only to relabel the variables as
$$A\leftrightarrow C,\quad p^\mu\leftrightarrow p^{\prime\mu},\quad
q^\mu\rightarrow q^{\prime\mu}.$$ 

The collision term can be written as
\be C[f_{A}]=-\int (-p^\mu q_\mu)f_A(\p)f_B(\q)W(\p,\q,\qp)\pi_+(\q)\pi_+(\qp).
\ee
Using the same decomposition that in the section \ref{spar62}, it follows
that its general form is
\be C[f_{A}]=-\sigma\int\Upsilon(m_A^2,m_B^2,p^\mu q_\mu,e^\mu,e^{\prime\mu})
(-p^\mu q_\mu)f_A(\p)f_B(\q)
\frac{d\Omega'}{4\pi}\pi_+(\q).\ee
Since $e^{\prime\mu}$ is arbitrary, the integration over $\Omega'$ is
straightforward and thus
\be C[f_{A}]=-\sigma\int\alpha(m_A^2,m_B^2,p^\mu q_\mu)
(-p^\mu q_\mu)f_A(\p)f_B(\q)\pi_+(\q).\ee We do now the ``Thomson''
approximation constiting in neglecting $(p^\mu q_\mu)$ compared with
$m_{A/B}^2$. In that limit the source term is given by
\be \epsilon_A^{(0)}=\sigma\alpha(m_A^2,m_B^2)\int(-p^\mu q_\mu)f_A(\p)
f_B(\q)\pi_+(\q)\pi_+(\p)=\sigma_{ABC}n_B^\mu n_{A\mu},\ee
which means that the reaction rate is proportional to the number of 
the two reactive constituant and to their relative speed.
Because of the symmetries we have $\epsilon_A=\epsilon_B=-\epsilon_C$.
Refinements can be included if there are any stoechiometric coefficients. \\

Let us now turn to the force,
\be F^{(0)\mu}_{\rightarrow A}=\sigma\alpha(m_A^2,m_B^2)\int
(-p^\nu q_\nu)p^\mu f_A(\p)f_B(\q)\pi_+(\q)\pi_+(\p)=\sigma_{ABC} 
n_{B\nu}T_A^{\mu\nu}.\ee
If we use the kinematic relations, we can check that
\be (1+\delta_{ABC})F^{(0)\mu}_{\rightarrow A}=
 (1-\delta_{ABC})F^{(0)\mu}_{\rightarrow B},\ee
which can be understood on two particular cases. If $m_A=m_B$ then 
$\delta_{ABC}=0$ and $F^{(0)\mu}_{\rightarrow A}=F^{(0)\mu}_{\rightarrow B}$,
i.e. each fluid undergoes the same force because of the
symmetry. If $m_A\gg m_B$ then $\delta_{ABC}\sim 0$ and
$F^{(0)\mu}_{\rightarrow A}\sim 0$, i.e the variation of impulsion
is very small for the fluid $A$ when it merges with the fluid $B$.

Furthermore, we have
\bea F^{(0)\mu}_{\rightarrow A}+F^{(0)\mu}_{\rightarrow B}&=&
\sigma\alpha(m_A^2,m_B^2)\int(-p^\mu q_\mu)f_A(\p)f_B(\q)
(p^\mu+q^\mu)\pi_+(\q)\pi_+(\p) \nonumber \\
&=&-\sigma\alpha(m_A^2,m_B^2)\int(-p^\mu q_\mu)f_A(\p)f_B(\q)
p{\prime^\mu}\pi_+(\pp)\equiv -F^\mu_{\rightarrow C},\eea
because of the kinematics relations and thus
\be 
F^{(0)\mu}0_{\rightarrow A}+F^{(0)\mu}_{\rightarrow
B}+F^{(0)\mu}_{\rightarrow C}=0.
\ee

We can give the correction to the source terms and the force
coming from the $(-p^\mu q_\mu)$-dependence in $\Upsilon$, which we develop
as in the section \ref{spar61}.
This implies a correction to the collision terme, $C_A^{(1)}$ say,
which is given by
\be C_A^{(1)}=-\frac{\sigma{\tilde\alpha}}{m_{AB}}\int 
p^\mu  q_\mu f_A(\p)f_B(\q)  \lambda\pi_+(\p),\ee
from which we can compute the correction to the
source term, $\delta\epsilon_A$ say,
\be \epsilon_A^{(1)}=-\frac{\sigma{\tilde\alpha}}{m_{AB}}
n_{B\mu} K_A^\mu,\label{eDE}, \quad{\rm whith}\quad
K_A^\mu=\int f_A(\x,\p)p^\mu p^\nu e_\nu\pi_+(\p).
\ee
The correction to the force, $F^{(1)\lambda}_{\rightarrow A}$,
is given by
\be
 \delta F^{(1)\mu}_{\rightarrow A}=-\frac{\sigma{\tilde\alpha}}
{m_{AB}}\int p^\mu p^\nu q_\nu \lambda f_A(\p)f_B(\q)\pi_+(\p) 
=-\frac{\sigma{\tilde\alpha}}{m_{AB}} J^\nu_AT_{B\mu\nu},\label{eDF}
\ee
where $J_A^\nu$ is defined in (\ref{eJA}).

\subsection{General case of annihilation}\label{spar73}

A general annihiliation can be written under the form
$$A(\p)+{\bar A}(\q)\rightarrow \gamma(\pp)+\gamma(\qp),$$
$A$ and ${\bar A}$ being two fermions.

An easy way to compute the source term and the force for such 
a mechanism is to use the work we have done on fusion and fission and assume
that the annihilation can be seen as
$$A(\p)+{\bar A}(\q)\rightarrow B\rightarrow \gamma(\pp)+\gamma(\qp),$$ 
with the decay time of $B$ being zero (i.e
$\tau_B\rightarrow 0$).

Hence the source terms are
\bea
\epsilon_A&=&\epsilon_{\bar A}=-\epsilon_\gamma/2 \nonumber \\
&=&\sigma_{A{\bar A}}n_A^\mu n_{{\bar A}\mu}. 
\eea

The force on $A$ and ${\bar A}$ is the same that the force computed
for fusion. The case of the photons is a little bit more
tedious. A solution will be to compute it from the integral 
(\ref{eFORCE}) as we
did in the former sections. We can however quote that we will
have to sum on two photons which are travelling in opposite directions
and which have the same energy. Thus, by symmetry, the total force
will be the same as if we had only one particle at the
end of the annihilation with the 4-momentum
$$P^\mu=p^{\prime\mu}+q^{\prime\mu},$$
and with the rest mass
$$M^2=-{\bf P}^2.$$
This is exactly the situation we have during a fusion and thus,
\bea
 F^\mu_{\rightarrow A}=F^\mu_{\rightarrow {\bar A}}=
 \sigma_{A{\bar A}}n_{(A}^\mu T_{{\bar A})\mu\nu}\qquad &
 F^\mu_{\rightarrow\gamma}=-2
 \sigma_{A{\bar A}}n_{(A}^\mu T_{{\bar A})\mu\nu}.
\eea

\subsection{Photons}\label{spar74}

Photons can be produced via Bremstrahlung which can be formally written as
$$ e(\p)\rightarrow e(\pp)+\gamma(\k).$$
If we follow Thorne \cite{THORNE}, the general collision term
for such a collision can be written as
\be C[f_\gamma]=E\chi_\gamma(E)\lb G_\gamma(E)-f_\gamma(\k)\rb n_e,\ee
where $E=-u_e^\mu k_\mu$ and $G_\gamma=\eta_\gamma/(E^3\chi_\gamma)$.
$\eta_\gamma$ and $\chi_\gamma$ are the standard emission and absorbtion
coefficients.
If we are in a local thermodynamical equilibrium, then 
Kirchoff's law holds and
\be G_\gamma=2\lp e^{E/T}-1\rp^{-1}.\ee
In general, the force on the photons can be written as
\bea F^\mu_{\rightarrow\gamma}&=&\int\frac{E^3}{m_e}
\chi_\gamma(E)G_\gamma(E)\lp u_e^\mu+e^\mu\rp n_e dEd\Omega \nonumber\\
&&-\int\frac{E^4}{m_e}\lp u_e^\mu+e^\mu\rp\chi_\gamma(E)f_\gamma 
(\k) n_e dEd\Omega.  
\eea
This can be computed only if we know the functions 
$\chi_\gamma(E)$ and $G_\gamma(E)$. However we can give its
general form, which must be
\be F^\mu_{\rightarrow\gamma}=\frac{1}{m_e}\lb
{\cal U}(\Theta_\gamma) n^\mu_e -q^\mu_\gamma{\cal V}(\Theta_\gamma)\rb,\ee
where ${\cal U}(\Theta)$ and ${\cal V}(\Theta)$
are two coefficients which depend on the photon temperature, $\Theta_\gamma$.
We cannot go further if we want to remain general. However the
form has the advantage to be covariant and flexible.

The situation is similar for the source term but, as for
the fusion which is in a way very similar, we can model it
as
\be \epsilon_\gamma=\frac{n_e}{\tau_e(\Theta_\gamma)}\quad
{\rm and}\quad\epsilon_e=0,\ee
where the rate of emission depends on the temperature.

If we have other fermions in the problem
then the force and the source term for the 
Bremstrahlung induced by these fermions will be 
alike, with coefficients ${\cal U}_f(\Theta)$, ${\cal V}_f(\Theta)$
and $\tau_f(\Theta_\gamma)$

\subsection{Recombination} \label{spar75}

A special case of interest in Cosmology is the
recombination of electrons and protons in hydrogen which
occurs at the last scattering surface \cite{CMB}. 
It is however different from the general case of fusion since the inverse
process is possible, the rate of each process depending on the
temperature.
$$e(\p)+p(\q)\rightleftharpoons H(\qp)+\gamma(\kp).$$
From our study of fusion and fission, it is clear that the
source term will take the general form (to lowest order),
\bea
\epsilon_H&=&\epsilon_\gamma=-\epsilon_e=-\epsilon_p \nonumber\\
&=&\sigma_{H\gamma}n_\gamma^\mu n_{H\mu}-\sigma_{ep}n_e^\mu n_{p\mu}
\eea
The two coefficients, $\sigma_{H\gamma}$ and $\sigma_{ep}$, are all
functions of the temperature.  Details on the ionisation and
recombination processes are needed to have their exact value. Examples
of such functions can be found in \cite{PEEBLES,ZEL}.

\subsection{Relation with quantum particle creation} \label{spar76}

We want to emphasize that,  
in our framework, even if there is particle creation via
unelastic collisions the total energy-momentum tensor is conserved.
The situation is then different from the one studied in cosmology
where the particles production arises from some quantum
process \cite{BD}.

Such phenomena were phenomenologically described by an
effective viscous pressure \cite{ZIM} and a microscopic justification
of such an approach was proposed in \cite{TRI}.
Their work is based on the introduction of a source term
in the Boltzmann equation beside the usual
collision term.  Unfortunately this source term gives also birth
to a force acting on the fluid (they only have one fluid).

If one wants to take into account such quantum creation of particles,
this can be achieved by using a linear coupling (see \cite{TRI}
section (3-1) $$C[f_i]=\lp-\frac{u^\mu p_\mu}{\tau(\x)}+\nu(\x)\rp
f_i(\x,\p),$$ where $u^\mu$ is, as usual, an arbitrary vector field
used to perform the splitting.

Using the same methods than before it will lead to a source term
and a force given by
$$\epsilon(\x)= -\frac{u^\mu n_\mu(\x)}{\tau(\x)}+\nu(\x)n(\x),$$
$$F^\mu(\x)= -\frac{u^\nu T^\mu_\nu(\x)}{\tau(\x)}+\nu(\x)n^\mu(\x).$$ 
In such a situation the global energy-momentum will not be conserved
(see \cite{TRI} for a detailed discussion of this problem). This is 
beyond the scope of this article.

\subsection{Recapitulation} \label{spar77}

Before we give some applications of our formalism, we will
sum up the different source terms and forces we have computed
either in the non-relativistic limit or in the Thomson limit.
We do not give the corrections than have been computed.
In fact, we computed them to show how one can give a general form. However,
as seen on equations (\ref{eF1A}), (\ref{eF1B}), 
(\ref{eFEP}), (\ref{eDE}) and (\ref{eDF}), 
they imply new macroscopic quantities.

For any applications, we need as input the cross sections
of all the collisions between the species we are considering and the
lifetime of all the unstable particles (i.e of
``$\sigma$", ``$\alpha$", ``$\beta$", ``$\tau$", ``${\cal U}$"
and ``${\cal V}$").\\

\begin{tabular}{|l||l|l|}
\hline 
 & $\epsilon$ & $F^\mu$ \\
\hline\hline
${\cal F}^{\mu\nu}$&0 &$F_B^\mu={\cal F}^{(0)\mu}_\alpha j^\alpha $\\
\hline
$A+B\rightarrow A+ B$ &0 &$F^\mu_{\rightarrow A}=
\sigma\lb \alpha n_{[B\nu}T^{\mu\nu}_{A]}+\alpha\rho_{[A} n^\nu_{B]}
\right.$\\
&&$\quad\quad+\left.\frac{\beta}{3m_B}T_{[B\nu}^\nu q^\mu_{A]} \rb$\\
\hline
$f+\gamma\rightarrow f+\gamma$&0 & $F^\mu_{\rightarrow\gamma}
  =\sigma_T\lp n_{f\mu}
  T^{\mu\nu}_\gamma+\rho_\gamma n_f^\nu\rp$\\ 
  && $F^\mu_{\rightarrow f}=-F^\mu_{\rightarrow\gamma}$\\
\hline
$A\rightarrow B+C$&$\epsilon_A =-\frac{n_A}{\tau_A}$&
 $ F^\mu_{\rightarrow A}=-\frac{n^\mu_A}{\tau_A}$ \\ 
 &$\epsilon_B=\epsilon_C=-\epsilon_A$& $F^\mu_{\rightarrow B/C}=
 \frac{1}{2}\lp 1\pm\delta_{ABC}\rp n^\nu_A/\tau_A$\\
\hline
$A+B\rightarrow C$&$\epsilon_A=\epsilon_B=\sigma_{ABC}n^\mu_An_{B\mu}$ & 
 $F^\mu_{\rightarrow A}=\sigma_{ABC}n_{B\nu} T^{\mu\nu}_A$\\ 
 & $\epsilon_C=-\epsilon_A$&$F^\mu_{\rightarrow C}=-2\sigma_{ABC}
 n_{\nu(B} T^{\mu\nu}_{A)}$\\
\hline
$A+{\bar A}\rightarrow \gamma+\gamma$&$\epsilon_A=\epsilon_{\bar A}= 
 \sigma_{A{\bar A}}n_A^\mu n_{{\bar A}\mu}$& 
 $F^\mu_{\rightarrow A}=F^\mu_{\rightarrow {\bar A}}=
 \sigma_{A{\bar A}}n_{(A}^\mu T_{{\bar A})\mu\nu}$\\ 
 &$\epsilon_\gamma=-2\epsilon_A$&$F^\mu_{\rightarrow \gamma}= -2
 F^\mu_{(\rightarrow A}+F^\mu_{\rightarrow{\bar A})}$\\
\hline
Bremstrahlung& 
              $\epsilon_\gamma=\frac{n_e}{\tau_e(\Theta)}$&
              $F^\mu_{\rightarrow\gamma}=\frac{1}{m_e}\lb
              {\cal U}(\Theta) n^\mu_e
              -q^\mu_\gamma{\cal V}(\Theta)\rb$\\
      &$\epsilon_e=0$&$F^\mu_{\rightarrow e}=-F^\mu_{\rightarrow\gamma}$\\
\hline             
$e+p\rightleftharpoons H+\gamma$& 
 $\epsilon_H=\sigma_{H\gamma}n_\gamma^\mu n_{H\mu}
 -\sigma_{ep}n_e^\mu n_{p\mu}$& \\
 &$\epsilon_e=\epsilon_p=-\epsilon_H=-\epsilon_\gamma$&\\
\hline
\end{tabular}
\\

%-------------------------------------------------------------------------- 
\section{Application to the theory of cosmological perturbations} \label{par8}
%-------------------------------------------------------------------------- 
\vskip 0.5cm

We give here a straightforward application of this general formalism
by considering the evolution of a gas of photons and electrons
interacting via Compton scattering in a Friedmann-Lema\^ \i tre universe
at linear order. The study of the formation and evolution
of the cosmic microwave background fluctuations begins usually
with the study of the radiative transport which is then integrated.
This has been done by many authors (e.g. \cite{KS} \cite{CMB})
in a perturbed Friedmann-Lema\^ \i tre universe. 

The metric of space-time is given by
\be
ds^2=a(\eta)^2\left(-(1+2A)d\eta^2 + 2D_iBdx^id\eta +(\gamma_{ij}
+h_{ij})dx^i dx^j\right)=\lp {\bar g}_{\mu\nu}+\delta g_{\mu\nu}\rp
dx^\mu dx^\nu,\label{MFL}
\ee
if we focus on scalar perturbations. $\gamma_{ij}$ is the metric
of the unperturbed $\lb t=constant \rb$-hypersurfaces, $D_i$
is the covariant derivative with respect to $\gamma_{ij}$,
$\eta$ is the conformal time, $a$ the scale factor and
a prime denotes a derivative with respect to $\eta$. 
We also introduce $h=h_{ij}\gamma^{ij}$.
Each fluid 
has an energy-momentum tensor given by
\be
T_{\mu\nu}=(P +\rho)u_\mu u_\nu + P ({\bar g}_{\mu\nu}+
\delta g_{\mu\nu})
+(\delta P + \delta\rho) u_\mu u_\nu + \delta P {\bar g}_{\mu\nu}
+2(P + \rho) u_{(\mu}\delta u_{\nu)} +a^2 P\Pi_{\mu\nu},
\ee
where $P$ and $\rho$ are the pressure and density of the fluid in the
background and $\delta\rho$, $\delta P$, $\delta u^\mu$ and $\Pi_{\mu\nu}$ are
respectively the density perturbation, the pressure perturbation, the
velocity perturbation and the anisotropic stress tensor. Since $u^\mu
u_\mu=-1$ then $\delta u_0=0$ and we can write $\delta u_k$ as
\be\delta u_k=a(D_kB+v_k).\ee 
We also introduce $c^2_m=dP_m/d\rho_m$, the sound speed of the
matter and $\delta\equiv \delta\rho/\rho$, the
density contrast. The law of evolution of the two
energy-momentum tensors are, according to equation (\ref{eFORCE})
\bea
\nabla_\nu T^{\mu\nu}_\gamma &=& F^\mu_{m\rightarrow\gamma},\label{CRAD}\\
\nabla_\nu T^{\mu\nu}_m &=& F^\mu_{\gamma\rightarrow m},\label{CMAT}
\eea
can be expanded to linear order. Since 
$u^\mu_mu_{\gamma\mu}=-1$ to first order, the force on the matter
(\ref{eFMPHOT}) reduces to
\be F^\mu_{m\rightarrow\gamma}= \frac{4}{3}\sigma_Tn_e\rho_\gamma
\lp u^\mu_m- u^\mu_\gamma\rp.\ee
Using the metric (\ref{MFL}) and the above expression of the force,
the equation of conservation (\ref{CRAD}-\ref{CMAT}) read
\begin{eqnarray}
\delta_\gamma'+\frac{4}{3}(\D v_\gamma+\frac{1}{2}h')&=&0 \\
\delta_m'+(1+\omega_m)\left[\D v_m +\frac{1}{2}h'\right]&=&0 \\    
(v_\gamma^i+B^i)'+D^iA+\frac{1}{4}D^i\delta_\gamma+\frac{1}{4}
D_j\Pi_\gamma^{ij}&=&\frac{4}{3}a\sigma_T n_e(v_b^i-v_\gamma^i) \\
(v_m+B)'+{\h}(1-3c^2_m)(v_m+B)-A&=&\frac{4}{3}\frac{\rho_\gamma}{\rho_m}
a\sigma_T n_e(v_\gamma-v_m), 
\end{eqnarray}
which is the usual result (see e.g. \cite{KS}). Let us stress that
since the Compton scattering is elastic, the number of photons
and electrons are conserved, which can be checked on the first
two equations.

If we now turn to the inelastic collision 
$$A+B\rightleftharpoons C+D$$
within the standard model of Cosmology, the equation of conservation
for $A$ reads
\be
\nabla_\mu n^\mu_A=-\sigma_{AB} n^\mu_A n_{B\mu}+\sigma_{CD} n^\mu_C n_{D\mu}.\ee
If we study a small deviation of $n^\mu_A$ from the equilibrium, then
\be\nabla_\mu n^\mu_A=-\sigma_{AB} n_{Beq\mu}\lp n^\mu_A-n^\mu_{Aeq}\rp,\ee
the subscript ``$eq$" meaning at equilibrium. In Friedmann-Lema\^ \i tre, this
becomes
\be {\dot n}_A =-3Hn_A-\gamma_A\lp n_A-n_{Aeq}\rp,\ee
with $\gamma_A=\sigma_{AB} n_{Beq}$. We recover the general
phenomenon of ``freezing". If $3H<\gamma_A$ then the system will 
relax toward the equilibrium, but, if $3H>\gamma_A$, the
system is decoupled since the interaction rate is not sufficient
to compensate the expansion rate. \\

Our formalism, being fully covariant, can be applied to
study the same problems in more general situations, for instance 
if we turn to inhomogeneous cosmologies.

%------------------------------------------------------------------------ 
\section{Conclusion}\label{par10}
%------------------------------------------------------------------------ 

We have developped a fully covariant framework for a system of
interacting fluids. We have computed general form of the source terms
$\epsilon_i$, and of the forces $F\mu_i$, related to the number flux
vectors $n^\mu_i$ and the energy-momentun tensors $T^{\mu\nu}_i$ by
$$\nabla_\mu n^\mu_i=\epsilon_i\quad{\rm and}\quad
\nabla_\nu T^{\mu\nu}_i=F^\mu_i.$$
We have considered the cases of elastic collisions (and the example
of the photon-electron scattering), of inelastic collisions (including
fission, fusion, bremstrahlung and recombination) and we have
included a possible magnetic force (see
table in section \ref{spar77} for a summary of the results for all
these cases). This computation first required
the modelisation of the collision term that enters the Boltzmann equation
and then the integration of this equation.

All the quantities have been computed for situations of cosmological
relevance (Compton scattering, recombination...) and can be used in a
wide range of redshift (e.g. $0<z<10^{12}$ if we just consider
photons-electrons and baryons).  Since we have the general force for
fluids constituted of massive particles, this formalism is also suited
for fluids of stars, galaxies...Moreover, thanks to the ``asymptotic
freedom", this formalism of interacting gases can be hoped to apply to
very dense systems, and thus during the early universe. However one
has to be careful in such an application and must check that the gaz
approximation holds.

This formalism is non perturbative, in the sense that we do not expand
the geometry around a background spacetime. It is of course
perturbative in the sense that we expand and integrate the Boltzmann
equation to a given order.  When expanded to first order (in the
perturbation of the metric), it reduces
to the equation of cosmological linear perturbations. But, our
formalism can be applied to study the dynamics of a system of gases in
a more general context. It will be useful, for example, to study the
microwave background in inhomogeneous cosmologies or compute the sound
speed of the photon-baryon fluid during the decoupling.

Let us stress that we do not assume anything on the fluids but the
fact that they are gases, i.e that they can be described by a
distribution function. When they are perfect fluids, the system of
equations is closed. Otherwise, we need either an equation of state or
an equation of evolution (e.g. for $\bar\Pi$) which can be obtained
from higher moments of the Boltzmann equation.

\section*{Acknowledgments} I am grateful to D. Langlois for
bringing this subject to my attention and for stimulating
discussions. I wish to thank B. Carter who helped me to clarify some
aspect of the subject, N. Deruelle for dedication in helping to
clarify the presentation as well as G. Faye, P. Peter,
S. Ratkovi\'c for their comments and G. and R. Uzan for their partial
financial support.

\end{document}